# Clinical Evaluation of Real-Time Optical-Tracking Navigation and Live Time-Intensity Curves to Provide Feedback During Blinded 4D Contrast-Enhanced Ultrasound Imaging

Ahmed El Kaffas Ph.D., Renhui Gong Ph.D., Samvardhini Sridharan, Rosa Maria Silveira Sigrist M.D., Isabelle Durot M.D., Jürgen K. Willmann M.D.*, and Dimitre Hristov Ph.D.

*Abstract* — **Current commercial matrix transducers for 3D Dynamic Contrast-Enhanced Ultrasound (3D DCE-US) do not display side-by-side B-mode and contrast-mode images when capturing volumetric data, thus leaving the operator with no position feedback during lengthy acquisitions. The purpose of this study is to investigate the use of transducer tracking to provide positioning feedback and to re-align images based on world coordinates to improve quantification. An interventional tracking system was developed in house using an infrared camera and a 3D-printed tracking target attached to a X6-1 matrix transducer. The system displays a virtual probe on a separate screen and allows to capture a reference position to provide operator feedback when no B-mode image is available. To test this set-up, five experienced operators were asked to locate an image landmark within a volunteer liver in B-mode images using the X6-1 connected to an EPIQ7 system (Philips, Bothell, WA). Operators were then asked to maintain the transducer position for 4 minutes under three feedback methods: i) B-mode, ii) display of real-time virtual transducer, iii) blind. The magnitude of displacement over the cine was computed as an estimate of the imaging position error. We also investigated whether transducer coordinates can be used to re-align images due to motion and improve contrast ultrasound perfusion repeatability. A total of 8 patient data were obtained under an IRB for this. Repeatability of perfusion parameters was assessed using the intra-class correlation coefficient. Results suggest that tracking can assist operators maintain a steady position during a lengthy acquisition. An average displacement of 3.75 mm with standard deviation (S.D.) of 3.31 mm was noted when using B-mode. With blinded acquisition, an average displacement of 4.58 mm (S.D. 2.65 mm) was noted. In contrast, the average displacement for tracking-feedback was comparable to B-mode at 3.48 mm (S.D. 0.8 mm). We also observed that perfusion parameters had better repeatability after re-alignment then without. To the best of our knowledge, this study is the first to demonstrate the feasibility of tracking for 3D DCE-US to provide feedback during lengthy scan sessions.**

*Index Terms* — **Contrast ultrasound, optical tracking, three-dimensional (3D) ultrasound,**

## I. INTRODUCTION

**U**ltrasound (US) is amongst the most widely accessible medical imaging modalities [REF to add]. In comparison to MRI or CT, it offers inexpensive bedside diagnostics without radiation and/or contrast restrictions related to kidney failure. As a result, functional ultrasound using 1D array transducers has been proposed as an ideal tool for longitudinal imaging applications such as cancer treatment monitoring to assess early tumor perfusion changes as surrogate for response (1).

An important limitation in longitudinal applications of ultrasound is positioning the transducer to obtain the same imaging plane for day-to-day comparison. In dynamic contrast-imaging protocols, the transducer must remain still in the imaging plane of choice. This is especially problematic in cancer imaging, where tumors tend to be highly heterogeneous, on a plane-to-plane basis [REF to add]. Thus, the operator-dependent nature of ultrasound can result in erroneous quantitative measurements and difficulties in navigating a transducer to different anatomical sites during time-sensitive scan sessions to capture different 2D planes (1).

Thus, the use of 3D DCE-US using matrix transducers for longitudinal imaging applications has been extensively suggested, and recently demonstrated as means to overcome the above-mentioned challenges by capturing whole-tumor perfusion simultaneously [REF to add]. However, several major challenges remain in the use of 3D DCE-US. One specifically relates to probe motion during acquisition, which can be imposed by either the operator or the patient. This can affect quantification and requires motion correction. However, unlike conventional 2D contrast modes in clinical systems, current commercially available 3D DCE-US modes do not display a low mechanical index B-mode and contrast-mode image side-by-side, thus leaving the operator with no positioning feedback during lengthy acquisitions to capture the dynamic contrast signal. In addition, without stable anatomical features from B-mode images, post-acquisition motion correction can be a significant challenge [REF to add]. As a result, the quality of lengthy acquisitions (> 30 seconds) to capture the temporal contrast dynamics during a bolus (2-4 min from wash-in to wash-out) or disruption-replenishment (> 4 min) acquisition can be compromised [REF to add]. Continuous positioning feedback such as available in 2D, would potentially improve the quality of the data for post-processing and quantification, by ensuring that the operator maintains a steady position throughout a scan.

We have recently developed an interventional acquisition system for 3D DCE-US imaging that aims to provide users with navigation feedback and temporal transducer coordinates to re-align the imaging planes in 4D for post-processing and quantification. The interventional system was also designed with disruption-replenishment (DR) imaging [REF] in mind, which has been shown to be more repeatable and quantitative over conventional bolus-based acquisitions [REF]. In brief, DR uses short high-power ultrasound pulses (disruptions; within diagnostic range) to momentarily burst microbubbles flowing at steady state. The rate of replenishment of microbubbles is then modeled to extract quantitative parameters (9). The advantages of DR is obviating the need to estimate the indicator input function (31). A key requirement for disruption-replenishment imaging is to ensure that perfusion has reached steady state in the tissue of interest. Given how heterogeneous tumors are, the time to steady-state is likely to vary from patient to patient. Currently, there is no method to confirm that steady-state has been reached live during acquisition. In this work, we also introduce a live TIC display for the operator to ensure that steady-state has been reached during a contrast infusion. This can help minimize the overall amount of contrast used, further decreasing potential risks linked to the procedure, as well length of the procedure.

The purpose of this study was to determine the feasibility of using optical tracking-based navigation assistance to provide positioning feedback in scenarios where no positioning feedback is provided. More specifically, transducer displacement during a lengthy acquisition was assessed under blind and B mode-guided scenarios and compared to navigation assisted positioning. We also evaluated in a subset of X patients if 4D image re-alignment improves quantification repeatability.

## II. METHODS

**Navigation System Overview**

The optical tracking system was designed and developed in house to facilitate transducer navigation, and to help maintain a specific imaging location over an anatomical site to overcome the lack of side-by-side contrast and B-mode image display in current commercial implementations of 3D DCE-US. The system displays on an LCD screen a live virtual probe that moves (translation and rotation) in sync with a real world optically tracked transducer. The system also captures reference coordinates and displays a second virtual probe in red at these coordinates to help operators maintain the same position and orientation of the transducer during lengthy scan sessions that can last up to 10 minutes.

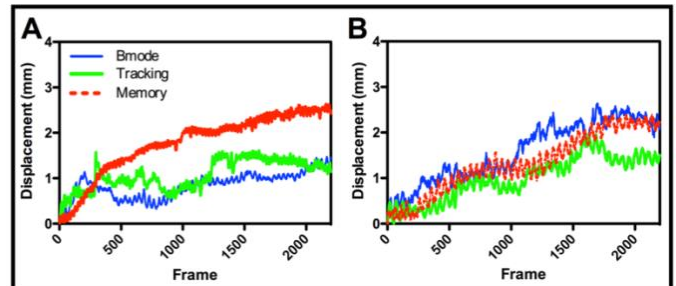

**Figure 2:** **A** and **B** are representative displacement of probe from reference position over 4 min for two different operators (operator 3 in A and 4 in B) shown under Bmode feedback, guidance feedback and memory. Note that A performs good with Bmode and guidance, while B performs best with guidance.

At the core of the platform is a modular research interventional workstation, which supports a Digital Navigation Link (DNL) for data transfer from a Philips EPQ7 (Philips, Bothell, WA) ultrasound system. Over a 1 Gb/s Ethernet connection, the DNL allows real-time transfer of 2D and 3D images from the scanners to the interventional workstation, which immediately displays them in real-time with user selectable layouts. Within the interventional workstation we have implemented and validated interfaces and modules to link ultrasound images to an external (world) coordinate system through optical cameras, which consist of a stereo pair of infrared (IR) cameras (Polaris, NDI, Canada) set-up over the patient bed (Figure 1A). Before each abdominal scan session, the cameras are oriented to face the lower-mid section of the abdomen using a laser pointer mounted at a mid-point between both cameras. The transducer was equipped with an in house 3D-printed contraption for holding four reflective spheres that are uniquely seen and tracked by the dual IR cameras (Polaris, NDI, Canada), and uniquely visible to the infrared cameras. The tool is set-up in such a way as to allow tracking translational and rotational movements of the transducer (3). For the work presented here, the tool was attached to an X6-1 (Philips, Bothell, WA) matrix transducer. The infrared cameras were connected to a PC running windows 7 and the MevisLab (Mevis Medical Solutions AG, Bremen, Germany) software package (Figures 1A-1E).

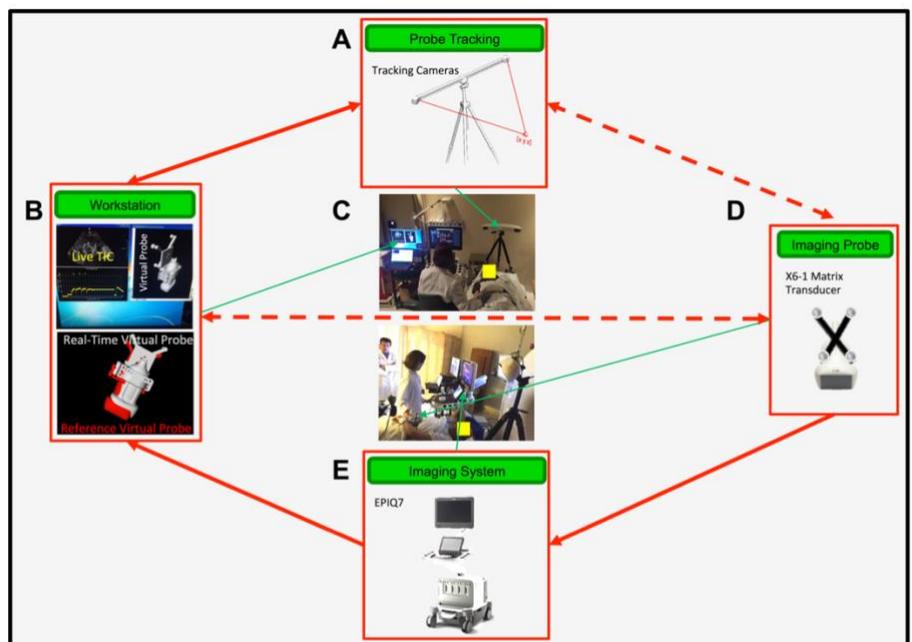

**Figure 1:** **A**, dual camera tracking system used to track transducer. **B**, interventional workstation used for acquisition of data (through DNL/Ethernet transfer) and visualization of live virtual probe. For live virtual probe based on tracking, an anatomical location is identified; a reference image is captured linked to a set of coordinates. A red still virtual reference probe (see lower picture; remains still) is then placed on the display to provide operator feedback. The user can then align the live virtual transducer (grey) with the red virtual transducer to reposition or maintain the same position during imaging. **C**, Image of set-up around a patient showing location of each instrument. **D**, X6-1 matrix transducer with 3D printed tracking markers used by the dual-cameras to determine location of the transducer. **E**, EPIQ7 system used to acquire data – the system is interfaced with interventional workstation to live-stream data via DNL/Ethernet to capture long (> 10 min) image sequences.

The MevisLab platform was used for real-time processing of incoming optical tracking data, acquisition of volumetric ultrasound data, and visualization of ultrasound data and virtual probes. A custom software was developed to handle these tasks. The software was developed using customized MevisLab modules written in C++ and Python. The user interface was implemented using MevisLab's custom MDL language with additional Python scripting. The software runs on the interventional computer workstation and communicates with the EPIQ7 ultrasound device and the Polaris optical tracking system. A 3D virtual model of the transducer with tracking tool was created in MevisLab with a specialized graphics unit, and linked to the streaming tracking coordinates through transformation matrices, which uniquely leverages the volumetric nature of 3D US (i.e. X6-1) (3). Calibration of of US image features is smaller than 1.5 mm (3).

All imaging evaluations in this work were carried out using a clinical EPIQ7 ultrasound system coupled to a clinical X6-1 transducer (Philips Healthcare, Andover, MA). The X6-1 2D matrix transducer has 9212 elements that steer the US beam in real-time using a micro-beam former located in the transducer head, and enables up to 90 deg by 90 deg wide volumes. It is designed for abdominal applications, with a frequency range of 1-6 MHz (center frequency, 3.2 MHz) (32). All contrast data was acquired in volumetric (3D) contrast-specific imaging (contrast-mode) with a low mechanical index (MI = 0.09) to allow non-destructive visualization of microbubbles. Disruption-replenishment used 3 flash frames with an MI of 0.77.

### Evaluation of Operator Performance with Navigation

Five experienced ultrasound operators were recruited to test our set-up in a clinical scenario reminiscent of a 3D DCE-US imaging session where long imaging sequences need to be acquired. To do so, we tested sonographers' ability to return to a reference position, or maintain the transducer in the same location with access to different sources of feedback. Each operator had at least 2 years of ultrasound imaging experience. Sonographers were first given up to 10 minutes to familiarize themselves with the tracking system and virtual probe display. Sonographers were then asked to locate an abdominal landmark using conventional B-mode images within a healthy volunteer. Once landmarks were located, operators were first asked to remove the transducer and return the exact same anatomical location on the abdomen. The repositioning error and time to return to the reference position were recorded as metrics. This was repeated for each operator twice. Operators were then asked to maintain the transducer position for 4 min under three feedback our method is based on the "hand-eye" technique (4). The method is fully automated with the use of data rejection based on sensor displacements, automatic registration of overlapping image regions, and a self-consistency error metric evaluated continuously during the calibration process. As a result, the uncertainty in spatial localization methods: i) conventional B-mode, ii) display of a real-time virtual probe and a reference virtual probe placed at the original location used to identify the abdominal landmark, iii) blind (no B-mode or virtual probe display). For the testing, sonographers used an X6-1 transducer with tracking attachment, connected to an EPIQ7 system (Philips, Bothell, WA).

**Table 1:** Re-positioning error (pos. in mm) and time to recovery (s) under different feedback methods.

| Opr. | Bmode | | | | Track | | | | Memory | | | |
|------|-------|------|-------|-------|-------|-------|-------|-------|--------|-------|-------|-------|
|      | Pos.  | Time | Pos.  | Time  | Pos.  | Time  | Pos.  | Time  | Pos.   | Time  | Pos.  | Time  |
| 1    | 1.83  | 14   | 5.32  | 16    | 12.19 | 26    | 0.39  | 23    | 22.46  | 14    | 36.39 | 9     |
| 2    | 7.57  | 4.00 | 3.53  | 7.00  | 7.25  | 13.00 | 1.20  | 11.00 | 28.42  | 14.00 | 20.68 | 7.00  |
| 3    | 0.87  | 15.00| 7.58  | 17.00 | 4.74  | 20.00 | 7.76  | 22.00 | 10.88  | 18.00 | 15.79 | 9.00  |
| 4    | 21.00 | 5.00 | 39.00 | 3.00  | 30.00 | 18.00 | 38.00 | 17.00 | 26.00  | 3.00  | 38.00 | 3.00  |
| 5    | 15.00 | 10.00| 2.04  | 9.00  | 9.64  | 33.00 | 2.97  | 50.00 | 17.51  | 6.00  | 8.84  | 10.00 |
| Mean | 9.25  | 9.60 | 11.49 | 10.40 | 12.76 | 22.00 | 10.06 | 24.60 | 21.05  | 11.00 | 23.94 | 7.60  |
| Stdev| 8.65  | 5.03 | 15.52 | 5.98  | 10.02 | 7.71  | 15.88 | 14.98 | 7.01   | 6.24  | 12.82 | 2.79  |

To quantify results, the magnitude of displacement of a center voxel in the image over the cine (acquisition sequence) was computed relative to the reference position as an estimate of the imaging position error. Histograms of displacements throughout the whole cine were also generated in order to examine the extent and directionality of displacement using histogram features such as the mean, median, standard deviation (S.D.) and skewness of the bins.

### Patient Inclusion for Evaluation in Clinical 3D DCE-US

Clinical 3D DCE-US patient data was obtained from an ongoing HIPPA compliant longitudinal prospective study approved by the Institutional Review Board of our institution and written consent was obtained from all participating patients. All patients were imaged with disruption-replenishment imaging with repeated pairs of disruption 'flash' events over an 8 minute infusion. The data was used to assess repeatability of quantitative 3D DCE-US parameters with and without tracking, and to evaluate the need for a live TIC tool. For this purpose, 8 adult (≥ 18 years old) patients with a total of 14 3D DCE-US scans were prospectively included. All patients with liver metastases ≥ 1 cm in diameter based on CT or MR imaging were considered eligible for our study and a clinical oncologist referred them to our study after introducing the study to the patients. Patients with documented anaphylactic or other severe reaction to any contrast media; pregnant or lactating patients; and patients with cardiac shunts or presence of severe pulmonary hypertension (all are contraindication for ultrasound contrast agent) were excluded. No patient was excluded due to exclusion criteria. Five patients were women (mean age, 54.5 years; range, 48 - 60 years) and 6 patients were men (mean age, 57.6 years; range, 47 - 68 years). Included patients had liver metastases originating from the following primary tumors: rectal adenocarcinoma (n=2); pancreatic adenocarcinoma (n=1); pancreatic neuroendocrine tumor (n=4); and colonic adenocarcinoma (n=4).

### Dynamic Contrast-Enhanced Ultrasound Imaging: Contrast Agent Administration, Acquisition and Quantification

Clinical contrast microbubbles (Definity; Latheus Medical Imaging, North Billerica, MA; FDA-approved for echocardiography and administered off-label for liver imaging in this study) were used. These microbubbles are perfluorobutane lipid microspheres with a mean diameter of 1.8 µm (range, 1-10 µm) (30). The DR DCE-US acquisition methods was used. Patients were infused for up to 3 min with a solution of 0.9 ml of the contrast agent mixed in 35.1 ml of saline at a constant rate of 0.08 ml/s (15) using a syringe pump (Medfusion 3500; Smiths Medical, Dublin Ohio) to reach steady state which was confirmed using an in-house developed time intensity curve (TIC) tool displaying TIC in real time on a separate monitor (Figure 3A). Two disruption-replenishments sequences (R1 and R2) were applied (2.5 minutes apart) with the contrast agent continuously infused to assess repeatability (Figure 3B). During the replenishment time (at destruction-replenishment DCE-US), all patients were asked to either hold their breath (for up to 30 seconds) or to breathe shallow (in patients unable to hold their breaths) to minimize motion artifacts.

Image analysis of 3D DCE-US was carried-out by one reader in random order using software developed in-house in MevisLab (MevisLab, Germany) and MATLAB (Mathworks, MA, USA). A volume-of-interest (VOI) was delineated by covering the entire liver lesion volume viewed on axial, sagittal, and coronal imaging planes. This VOI was subsequently used to generate TICs for the quantification of perfusion as described below. Post-processing and analysis steps for DCE-US following VOI selection consisted of the following steps: (1) linearization of the US image voxel values in contrast-mode images using a transformation function and a compression parameter provided by the equipment manufacturer (34); (2) extraction of TIC proportional to contrast concentration from the average signal intensity in the VOI; (3) standardized monoexponential fitting to VOI average intensity (TIC) using standard quantitative models in Matlab (9,35). The custom analysis software permitted the selection of VOIs around the lesion, and the application of dynamic enhancement models that could not otherwise be applied in 3D by using commercial software. Monoexponential curve fitting was performed from the first frame after the disruption event for a minimum of 1 min for DR DCE. The following parameters were extracted: relative blood volume (rBV), and relative blood flow (rBF) (35). The rBV is proportional to blood volume; the rBF is proportional to flow/perfusion rate (Figure 2A and 2B).

**Table 2:** Mean transducer displacement from reference position (in mm) under different feedback methods, during a 4min cine.

| Operator | Bmode | | | | Track | | | | Memory | | | |
|---|---|---|---|---|---|---|---|---|---|---|---|---|
| | mean | std | skw | median | mean | std | skw | median | mean | std | skw | median |
| 1 | 2.11 | 3.55 | 4.90 | 1.30 | 5.60 | 4.70 | 1.15 | 3.50 | 1.00 | 2.60 | 19.80 | 0.80 |
| 2 | 6.87 | 6.40 | -0.80 | 11.00 | 1.60 | 1.50 | 10.64 | 1.35 | 6.30 | 17.80 | 15.80 | 5.00 |
| 3 | 1.85 | 4.80 | 12.00 | 1.03 | 1.10 | 1.40 | 20.60 | 1.00 | 2.10 | 3.70 | 14.60 | 2.10 |
| 4 | 2.80 | 5.70 | 21.00 | 2.40 | 3.10 | 10.50 | 8.00 | 1.60 | 8.50 | 21.70 | 3.00 | 2.10 |
| 5 | 1.15 | 3.90 | 20.50 | 0.90 | 2.60 | 6.60 | 8.90 | 1.50 | 2.70 | 10.30 | 7.00 | 1.30 |
| 6 | 1.76 | 10.70 | 9.50 | 0.70 | 1.10 | 0.50 | 0.70 | 1.00 | 1.00 | 0.60 | 0.30 | 0.80 |
| Mean | **2.76** | **5.84** | **11.18** | **2.89** | **2.52** | **4.20** | **8.33** | **1.66** | **3.60** | **9.45** | **10.08** | **2.02** |

## III. STATISTICAL ANALYSIS

To measure repeatability of DCE-US quantitative parameters from DR DCE-US data sets, pairs of log-transformed measurements were assessed by intraclass correlation coefficient (ICC) from a random-effects model, with random effects of scan session (SS) nested within patient. Log-transformation was applied to make the data normally distributed for standard statistical analysis. The 95% confidence intervals (CI) was also calculated for each ICC. ICC of 0-0.20 indicated no agreement; ICC of 0.21-0.40, poor agreement; ICC of 0.41-0.60, moderate agreement; ICC of 0.61-0.80, good agreement; and ICC greater than 0.80, excellent agreement (36). All statistical analyses were performed using Stata Release 14.1 (StataCorp LP, College Station, TX). Statistical significance was fulfilled at P < 0.05.

## IV. RESULTS

**Tracking-Assisted Positioning Assessment**

Results suggest that our tracking system can assist operators return to, or maintain a transducer position during a lengthy acquisition session, such as those commonly employed in contrast-enhanced ultrasound, by providing positioning feedback through the use of virtual transducer display. Overall, sonographers found the use of a virtual display of the probe for positioning feedback clinically feasible, but required familiarization with the system before the start of system testing sessions.

Table 1 shows repositioning error and time to reposition for each sonographer repeated trials, specified for each of the feedback methods. Note that while repositioning under Bmode feedback offered the most rapid and accurate results, tracking-assisted repositioning offered slightly worst results. No feedback resulted in large repositioning errors.

The average relative displacement throughout the acquisition sequence was used as a metric to assess the three different positioning feedback methods. Figure 1F shows a representative B-mode of an abdominal landmark in the liver used to assess our system. Figure 2 exhibit the magnitude of the displacement of a center voxel and associated histogram over the 4 min acquisition for each of the feedback conditions, for sonographer 1 and X. Table 2 is a summary of all histogram features (mean, median, S.D. and skewness) for each of the sonographers. An average displacement of 3.75 mm with standard deviation (S.D.) of 3.31 mm and displacement histogram skewness of -0.18 was noted when using B-mode as feedback. When blinded (no B-mode or virtual probe display), an average displacement of 4.58 mm (S.D. 2.65 mm; skewness 6.19) was noted. In contrast, the average displacement for tracking-feedback using a virtual probe was comparable to that from B-mode at 3.48 mm, with a smaller standard deviation and less skewness of the displacement histogram (S.D. 0.8 mm; skewness 0.09). One operator performed better with tracking than B-mode. When blinded (no B-mode or virtual probe display), an average displacement of 4.58 mm (S.D. 2.65 mm; skewness 6.19) was noted. In contrast, the average displacement for tracking-feedback using a virtual probe was comparable to that from B-mode at 3.48 mm, with a smaller standard deviation and less skewness of the displacement histogram (S.D. 0.8 mm; skewness 0.09). One operator performed better with tracking than B-mode; one operator performed better blinded than with tracking and B-Mode. The use of tracking for positioning feedback resulted in more consistent (smaller S.D.) results, and a mean displacement that was comparable to B-mode- based positioning feedback.

**Clinical Use of Interventional System**

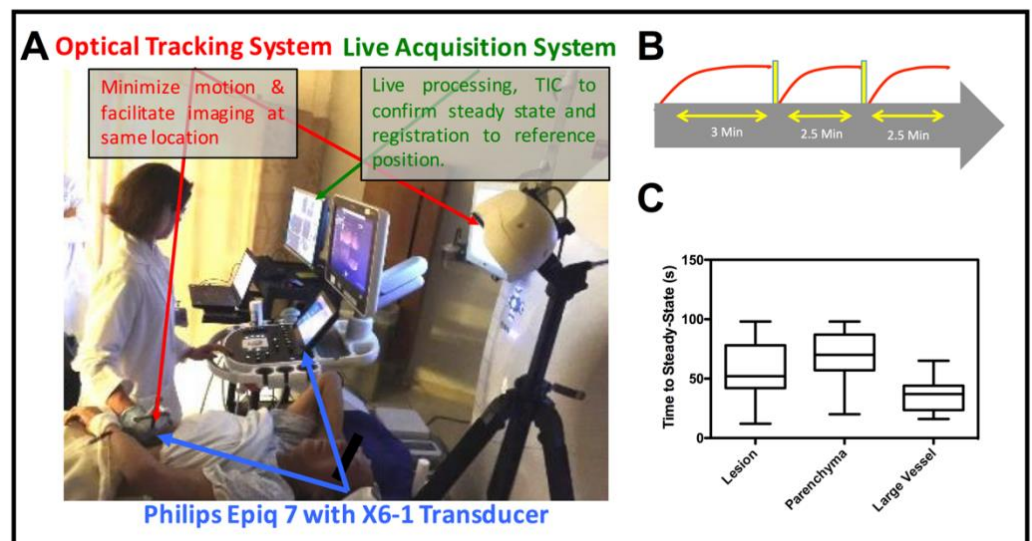

**Figure 3: A** Picture of clinical set-up of EPIQ7 interfaced with the interventional unit. **B** Disruption-replenishment imaging timeline. **C** Time to steady state shown as a box and whisker plot for all patients enrolled at different anatomical sites available in the 4D volume, within the liver.

Overall, sonographers using the interventional system found its use feasible and, after some familiarization as per above, they qualitatively found that it can help improve positioning and maintaining a position during blind scans, especially during contrast sequences after contrast clears through disruption.

In order to confirm the need for a live TIC tool, using the complete infusion acquisition sequence, we calculated the time it takes for the contrast intensity to reach steady state in 3 different tissue: the liver lesion, the liver parenchyma, and the portal vein. The time to steady state is presented in Figure 3C as a box and whisker plot for all patient scans. Note the large variability in the time-to-steady-state, specifically in the liver and lesion parenchyma, which could be affected by general patient physiology and especially by the histology of the primary tumour of the metastasis that would differ on a patient-by-patient basis. Overall, the live TIC tool was helpful in determining timing of disruption events.

**Improvement in Parameter Repeatability**

In order to determine if the tracking coordinates can help compensate for some operator/patient motion by re-aligning the imaging frames, we measured the repeatability of perfusion parameters obtained consecutively for the same VOI using the DR method. ICC results for the rBF and rBV repeatability measurements are presented in Table 3, with and without tracking re-alignment.

**Table 3:** ICC repeatability (with 95% CI) of DR parameters.

|  | No Tracking | Tracking |
|---|---|---|
| **rBF** | 0.72 (0.68,0.73) | 0.89 (0.88,0.90) |
| **rBV** | 0.89 (0.89,0.089) | 0.99 (0.97,0.99) |

## V. DISCUSSION

To the best of our knowledge, this study is the first study to demonstrate the feasibility of tracking for 3D DCE-US to provide feedback during lengthy scan sessions. Our results are in general support of the use of tracking in conditions where side-by-side B-mode imaging is not available, such as in current commercial implementations of 3D DCE-US. These also suggest that tracking and the use of a virtual probe could yield more consistent results (smaller displacement S.D.). It is also possible that if operators were given more extensive training and familiarization time with the tracking system, that their performance could be further improved.

An advanced ultrasound transducer positioning system could provide positioning feedback during imaging and minimize quantitative errors arising from the operator-dependent nature of ultrasound. Several manufacturers have developed real-time tracking and positioning systems to address this issue with some success in specific medical applications, but generally limited by: usability, errors due to interference with positioning signal (electromagnetic), and difficulty to visualize transducer position relative to a reference position. We have developed an advanced positioning system that attempts to overcome some of these limitations. The in house-developed system uses a custom live display of a virtual transducer for positioning guidance, and is designed for use with current implementations of 3D DCE-US imaging. In integrated this live positioning/tracking system into an interventional cart that also included a live TIC display to help guide DR DCE-US imaging to identify contrast steady state, and demonstrated the need for such a tool in live patients imaged with 3D DCE-US. The live TIC tool specifically offers an avenue to individualize timing of imaging during contrast-infusion, minimizing the contrast used on a patient-by-patient basis to minimize risks linked to procedure, all while maintaining or improving the quantification of DR imaging sequences.

It is anticipated that with further development, our virtual tracking system could enable longitudinal re-positioning in treatment monitoring applications if fiducial markers are identified or placed on the patient, and an IR camera can be permanently fixed in the ultrasound scan room. In addition, several modern mobile systems could help omit the need for much of the equipment-heavy optical tracking systems [REF to add]. Finally, as demonstrated, the use of tracking could facilitate volumetric motion correction for improvement quantification, and with the small optical tracking system uncertainty (3), enhanced spatial alignment could be used to improve quantification and voxel-based analysis in 3D.

An important limitation to our study is that we did not take into account internal shifts of the identified anatomical landmark resulting from cardiac or respiratory motion. In the future, acquisition of volumetric B-mode data using a matrix transducer under each of the feedback conditions used in this study could be used to assess the displacement of the identified landmark within the image. This could be done by registering each consecutive image within the cine acquisition sequence, and extracting the root-mean-square-error of the registration, thus telling of an operators' ability to not only maintain the same transducer location, but to also maintain the same image. With volumetric imaging capabilities available on modern matrix transducers such as the X6-1, both in plane and out of plane motion could be assessed and used to further improve our system in conjunction with the transducer tracking coordinates.

In summary, our study demonstrates the feasibility of using a tracking system along with a virtual probe display to overcome current limitations with commercial implementations of 3D DCE-US, where side-by-side imaging is not available. Our results also suggest that with additional operator training and further development of our system that it may become superior to conventional B-mode imaging for transducer positioning.

## VI. ACKNOWLEDGEMENTS

We thank Vijay Shamdasani, PhD (Philips Healthcare, Bothell, Wash), for providing the clinical US system and transducer. We would also like to acknowledge Dr. Peter Burns from the Sunnybrook Research Institute/University of Toronto for discussions.